\documentclass{elsarticle}
\journal{arXiv}
\usepackage{a4wide}
\usepackage{mathtools}      
\usepackage{amstext, amsmath, amssymb}
\usepackage{graphicx}
\usepackage{latexsym}
\usepackage{booktabs}
\usepackage{url}
\usepackage{import}
\usepackage{fancyvrb}
\usepackage{stmaryrd}
\usepackage{pdfsync}
\usepackage{algorithm}
\usepackage{tikz}
\usepackage{pgfplots}
\usepackage{pgfplotstable}
\usepackage{tabularx}
\usepackage{todonotes}
\usepackage{subfigure}
\UseRawInputEncoding
\usepackage{array}
\usepackage{verbatim}

\usepackage[plainpages=false]{hyperref}
\hypersetup{
		bookmarksopen=true,
		bookmarksnumbered=true,
		pdfborder={0 0 0},
		colorlinks=true,
		linkcolor=blue,
		urlcolor=blue	
              }
              
\DefineVerbatimEnvironment{code}{Verbatim}{frame=single,rulecolor=\color{blue}}

\numberwithin{equation}{section}
\numberwithin{figure}{section}
\numberwithin{table}{section}

\newdefinition{remark}[theorem]{Remark}
\newproof{proof}{Proof}

\numberwithin{equation}{section}

\journal{arXiv}

\begin{document}

\begin{frontmatter}
  
  \title{An Automated and Efficient Aerodynamic Design and Analysis Framework Integrated to PANAIR}

\author[khas]{Tahura Shahid}
\ead{tahura.shahid@khas.edu.tr}

\author[khas]{Ceren G\"urkan \corref{cor1}}
\ead{ceren.gurkan@khas.edu.tr}
\cortext[cor1]{Corresponding author}

\address[khas]{Department of Civil Engineering, Kadir Has University, TR-34083 Cibali, Istanbul, Turkey}

\begin{abstract} 
Aircraft design is an iterative process that requires an estimation of aerodynamic characteristics including drag and lift coefficients, stall behavior, velocity, and pressure profiles repeatedly, especially in the conceptual design phase. PanAir is a high-order aerodynamic panel method-based algorithm developed as a part of the Public Domain Aeronautical Software program mostly with NASA sponsorship. It is based upon potential flow theory and is used to numerically compute lift, induced drag, and moment coefficients of the aircraft in both subsonic and supersonic flight regimes. Quoting from developers \textit{it is the most versatile and accurate of all the linear theory panel codes}. PanAir is a classical software that requires geometric data as an input in the form of a PanAir-compatible format; however, commonly used Computer-Aided Design Software packages no longer conform to the PanAir input format. Likewise, PanAir produces its output in a PanAir-specific output file which is not compatible with commonly used visualization software. The input geometry required by PanAir and its output, therefore, involves significant manual pre- and post-processing using intermediary software. The work proposed here is an automated pre and post-processor to be used together with PanAir. With the environment proposed in this work, manipulation of input and output data using several intermediary software to and from PanAir is bypassed successfully. The proposed environment is validated over a Cessna 210 aircraft geometry with a modified NLF (1)-0414 airfoil. The aircraft is numerically analyzed using PanAir together with the proposed environment and results are compared with full-scale wind tunnel data.  
\end{abstract}


\end{frontmatter}

\section{Introduction}

Lately, the aerospace industry has experienced a shift toward a more integrated approach for aircraft design. This integrated environment permits a comprehensive analysis of a aircraft's aerodynamic performance. It typically includes an aerodynamic tool, a user-friendly interface, and a data management tool.
The aerodynamic tool is used to predict the aerodynamic performance based on numerical simulations of the fluid flow around the aircraft; a user-friendly interface allows engineers to set and run simulations efficiently and visualize the results; the data management tool is used to store all the aerodynamic data. This holistic approach to aircraft design enables engineers to save time and make informed trade-off decisions between various design parameters. Also, this integration provides engineers with a powerful toolset for aircraft design, allowing them to evaluate the impact of design changes on certain crucial aspects such as aerodynamics, stability, control, and weight in real time.

One such advanced computational tool is PanAir \cite{PanAir} \cite{PanAir1}. Since its inception, PanAir has become an indispensable tool in the aerospace industry. Still recently, giants of aircraft design and manufacturing like NASA, Boeing, Airbus, and Lockheed Martin have been working with PanAir on cutting-edge technology development. It enables the engineers to simulate and accurately predict the aerodynamics of airfoils, wings, and overall aircraft configuration. One of the critical advantages of PanAir is its ability to predict aerodynamic loads on aircraft accurately. This includes lift, drag, pitching moment, and rolling moment, which are essential for engineers at design stage. Another reason for the software's widespread use is its ability to handle a sufficiently large number of boundary conditions, including the boundary condition over propellers, engine nacelle, and other auxiliary components such as antennas and drop tanks\cite{CavallaroDemasi2016}. With PanAir, aircraft design optimization can also be handled \cite{WangWanLiuYang2018}. Moreover, PanAir can be used to analyze different design configurations, which helps to evaluate the impact of changes to the airframe, such as wing sweep, taper ratio, and winglets. This leads to the selection of best design for a given set of requirements. Finally with PanAir, the position of engines, fuel tanks, and other components can be optimized; helping to achieve the best balance between performance and efficiency \cite{CavallaroDemasi2016}. However, the software does not account for transonic flows, viscous flows, or flows with adverse pressure gradients \cite{PanAirManuel}.

\subsection{Background}\label{background}

PanAir is a numerical algorithm that solves potential flow equations, using the high-order panel method. The panel method is a  method used in fluid dynamics to analyze the flow around an object by only discretizing the boundary of the domain. It is based on the concept of dividing the surface of the object into a set of discrete panels and then using mathematical methods to calculate the flow characteristics at each panel.

The panel method assumes that the flow around the object is irrotational, meaning that the fluid has no vorticity. This assumption simplifies the complex 3D Navier-Stokes fluid flow equation into a simple Laplace equation which relates the velocity potential (a scalar quantity that describes the flow) to the distribution of singularities, such as sources/ sinks, vortex, or doublet on the surface of the object. These singularities (depending on the panel method type) are placed at the center of each panel, and the magnitude of these singularities over each panel is calculated using the imposed boundary condition. The magnitude of these singularities gives the potential function, which can be used to compute the induced velocity at each panel using a system of equations.

Once the flow characteristics are calculated at each panel, they can be used to determine performance parameters, such as lift, drag, and pressure distribution. While the panel method has certain limitations and assumptions (as elaborately described in PanAir Manual \cite{PanAir_Manual}), it remains a valuable tool for analyzing a wide range of fluid dynamics problems.

\begin{figure}[H]
  \begin{center}
    \includegraphics[width=0.90\textwidth]{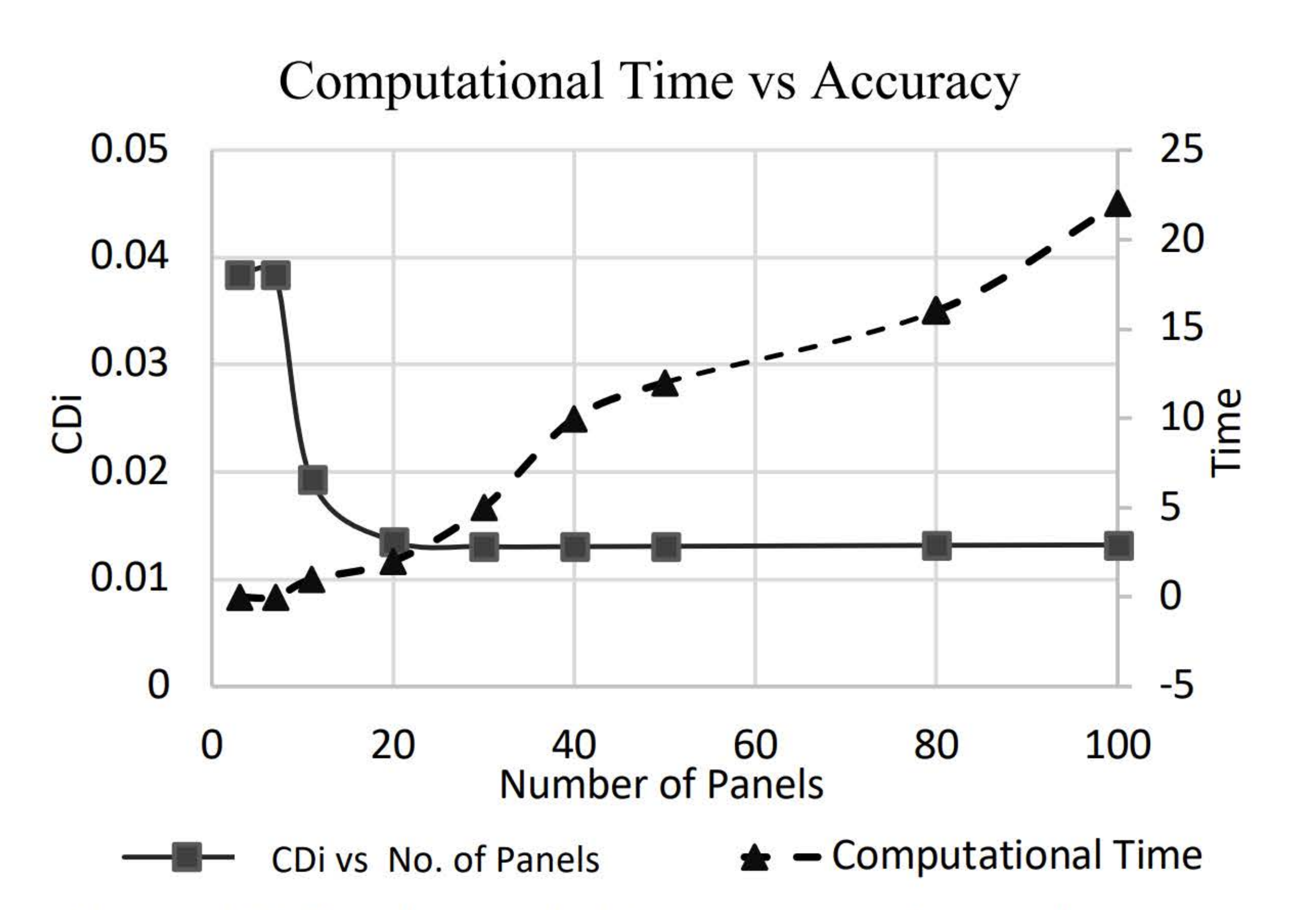}
\end{center}
\caption{The computational time Vs Number of panels used and accuracy \cite{ShahidSajjad2020}}
  \label{fig:my}
\end{figure}

There are various types of panel methods but the most popular one is the higher-order PanAir. It is called higher order because of the quadratic distribution of the doublet. The major advantage of the higher-order panel method is that its accuracy is independent of panel size and aircraft configuration \cite{ShahidSajjad2020}. Shahid et al. drew a significant conclusion by performing mesh independence, over a trapezoidal wing platform, using PanAir and then generalized it for all aircraft configurations. It is observed that with sufficient number of panels, accurate results can be achieved within an acceptable error range. In the study, the number of panels is varied from 0 to 100, and a comparison is made between the computational time, the number of panels, and the induced drag. It was observed that increasing the panel number over 20 only slightly changed the induced drag but increased the computational time significantly, see Figure \ref{fig:my}.

Typically, PanAir requires detailed, paneled geometry descriptions expressed as coordinates of quadrilateral panels. To accomplish this, the geometry of interest is first discretized into panels, and the resulting node-point coordinates are extensively manipulated to conform to the right-hand rule. The right-hand rule is a convention used in the panel method to determine the orientation of the normal vector at each panel. By convention, the direction of the normal vector should be perpendicular to the panel surface and is crucial for calculating the flow characteristics. It is determined using the right-hand rule, which states that if the fingers of the right hand are curled in the direction of the panel-wise singularity strength, called circulation, the thumb should point in the direction of the normal vector. The use of the right-hand rule ensures that the normal vectors are consistently oriented across all panels, which is necessary for the accurate modeling of the flow.

\begin{figure}
  \begin{center}
    \includegraphics[width=0.90\textwidth]{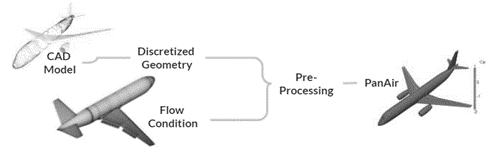}
\end{center}
\caption{The PanAir input work flow chart}
  \label{fig:PanAirWorkflow1}
\end{figure}

This data is combined with flow-field information that includes Mach number and angle of attack and then manually translated into PanAir-readable input (.inp) file. These pre-processing steps require significant man-hours and computational resources that makes PanAir not a viable option in time-effective conceptual design process. Furthermore, the PanAir produced output file (.agps) once more requires manual and time-consuming translation into graphical format for flow visualization over the aircraft surface in any post processing tool. The entire process can take months to accurately model and simulate even a simple aircraft geometry. The flow chart in Figure \ref{fig:PanAirWorkflow1} represents the input workflow of PanAir.    
\newpage
\subsection{Earlier work}
PanAir was developed by Boeing in 1970s in collaboration with Ames Research Centre NASA and United States Airforce \cite{PanAir1}. A pilot version was made available by Boeing online in 1990s; refereed as A502. After 50 years, the code is still used extensively in different aspects of aircraft design such as for sonic boom prediction \cite{Tekaslan} and minimization \cite{Demiroglu}, fast analysis  of stealth aircraft \cite{Sepulveda}\cite{Kumser}, static aeroelasticity prediction \cite{Jovanov}, design optimization \cite{Sun} and many other. 

Bolander and Hunsaker \cite{Bolander} utilized Panair and presented work on designing supersonic aircraft using low-fidelity tools. They proposed a novel method for splicing near-field pressure signatures using Panair to predict the overall pressure signature of the supersonic aircraft. In this research, the computational efficiency of PanAir was exploited to create the design space and then trained a neural network to learn the relation between near-field and overall pressure signature. The method was demonstrated on a delta-wing aircraft, and the results showed that the spliced pressure signature estimation matches well with the high-fidelity simulation while the computational cost and time were considerably reduced. 

Similar work was done by Tekaslan et al. in \cite{Tekaslan}, where they presented a new approach for predicting sonic boom noise from supersonic aircraft while accounting for uncertainty using PanAir. The results depict that their UQ (Uncertainty Quantification) framework was highly effective in predicting sonic boom noise while accounting for uncertainties in aerodynamics, acoustics, and structure. The results were also compared with the contemporary Monte Carlo approach and stated that their approach using PanAir is more efficient in handling multidisciplinary problems.

A much-related study was done by Demiroglu et al \cite{Demiroglu} for minimizing sonic boom noise of a supersonic aircraft using parametric wing shape design. A multi-fidelity optimization framework to optimize supersonic aircraft wings with the objective of minimizing sonic boom noise was proposed. PanAir proved to be very efficient for full exploration of the design space, and high-fidelity tools provided an accurate prediction of sonic boom noise. The results showed that there was a significant reduction in the sonic boom noise, and the aerodynamic efficiency was improved. 

A new approach was introduced in \cite{Kumser} for aerodynamic analysis and design of jet aircraft using PanAir. In this work, researchers performed a comparative study using PanAir and experimental data to validate the accuracy of their approach using an aircraft similar to F-16. The study was performed over various Mach Numbers and angles of attack. Later, the group modified the aircraft’s wing and fuselage to improve aerodynamic efficiency. The results showed a significant improvement in aerodynamic performance compared with the original design. 

Due to PanAir's ever-increasing use for aerodynamic analysis, there is a clear need to make it automatic and integrated with the intermediary tools required for pre and post-processing. The use of PanAir is very tedious and time-consuming without that integration. Although PanAir is available online as a complete package with a geometry creator, a translator (named MakeWGS) and a pre-processing file called Panin, these pre-processors  are either limited or need manual manipulations. 
The MakeWGS executible file creates a very simple geometry by taking a few geometric parameters, creating quadrilateral panels on the surface, and translating them into NASA Langley Wireframe Geometry Standard (LaWGS) format \cite{Holt}. This then can be a direct input file to the Panin together with the flight data (auxiliary) file. However, MakeWGS constraints the geometric data and fail to model complex geometries \cite{Schmidt}. Therefore, the researchers use alternative CAD software to model complex configurations. These softwares can capture even the finest geometric detail such as a winglet, pylon and nozzles \cite{Rodriguez}, and drop tanks \cite{LeTourneau}. Thereby using an alternative CAD software instead of MakeWGS for geometry discretization highly improves the accuracy of the analysis. However, geometric discretization created using alternative CAD software should be translated into LaWGS format (input format required by PanAir), which necessitates using other intermediary software. Hence, despite the unprecedented benefit of improved accuracy, an added step for input file translation prolongs the computational time.  

The need to develop an integrated framework for PanAir gained popularity and many researchers and engineers took part in this quest. The research focused on making the pre and post-processing of the PanAir automatic, user-friendly, and easy to handle. Schmidt and Brunswig \cite{Schmidt} utilized PanAir to analyze the flow around AC20.30 blended wing body. Being aware of the limitation of the pre and post-processors included with PanAir, they developed their own strategy by using Gambit for discretization and PanView, which is a Matlab-built post-processor \cite{Schmidt}, for visualizing the flow over the geometry. As a result, the time required for the analysis was reduced from several months to a few hours. Still, it was not an optimal solution as other software or tools were still utilized to resolve abutment issues, to generate wing tips, and to maintain flow tangency. These programs were not integrated into a single unit and thus required manual interpretation by the user. 

Similarly, Mehdi and Masud \cite{Mehdi} devised a scheme to operate PanAir with Gambit for creating panels. The geometric input file obtained from Gambit is then made compatible with the PanAir input format using Excel, and the results are visualized through Tecplot. However, these tasks are not interlinked through an integrated environment, so this workflow required considerable time and manpower. Tarkian and Tessier \cite{TarkianTessier2008} took a step further and, for the first time, introduced a semi-automatic in-house developed integrated framework for PanAir with Catia. The interface was developed specifically for typical civil jet aircraft, where modeling and paneling were both performed in Catia V5. The generated geometric input was then fed into a novel translation program that converts the geometric input into PanAir compatible format. Also, this translation program rearranged the number of panels and resolved the abutment issues. The only issue with this unique interface was that it still involves manual geometry manipulation and is semi-automatic.  The two software, PanAir, and Catia were not linked, and the user must perform operations between the two manually. Post-processing was not considered in this setup.

Amadori, Jouannet, and Krus \cite{AmadoriJouannetKrus2006}  developed a framework with Matlab and PanAir for aerodynamic shape optimization. The algorithm was implemented in Matlab, which had all the geometric data obtained from Catia and then translated into PanAir-compatible format through another Matlab script. An entire framework developed in Matlab linked the pre-processor to the PanAir. The framework included only the pre-processing and data translation in the required format and did not cover post-processing. An alternative integrated design framework was built by Doyle et al. \cite{Doyle}. The framework included integrating high, mid, and low-fidelity aerodynamic tools for aircraft design. The high-fidelity tools included CART3D and Usm3D; low-fidelity included PanAir and S/HABP; the mid-fidelity was just an interpolation of high and low-fidelity data sets. For the discretization of geometry, an in-house developed GMAP tool was used. However, there were limitations regarding the geometry modeling. The geometry had to be manually created and abutted and it is stated that the difficulties were faced while creating the model for PANAIR. Yet another tool used in this work with PanAir was S/HABP but only simple geometries were able to handle it. Moreover, Doyle and his group could depict the data obtained by PanAir through plot3D \cite{Doyle}.

The most prominent work in the front of an integrated pre and post-processor environment to be used with PanAir was proposed by Cavallaro \cite{RaunoCavallaro} regarding the development of a Graphical User Interface with a surface modeler (panel creator) for PanAir. The geometric parameters were taken as input from the user; thus, this program gave agility and robustness to the aircraft design process. However, it has limitations, including difficulty importing a Catia file, engine modeling or issues in generating quadrilateral interconnected panels \cite{CavallaroDemasi2016}.

\begin{table}[ht]
    \centering
    \caption{Summary of related work}
    \resizebox{\textwidth}{!}{
    \begin{tabular}{|c|c|c|c|c|}
    \toprule
         Researchers & Geometry Modeller & Parameterization & Visualization & Application \\
    \midrule
         Schmidt and Brunswig \cite{Schmidt} & Catia & Gambit & PanView & AC20.30 BWB \\
    
         Mehdi and Masud \cite{Mehdi} & Catia/ Gambit & Gambit & Tecplot	& JF-17 Fighter aircraft \\
    
         Tarkian and Tessier \cite{TarkianTessier2008} & Catia & Catia & None & Boeing 777 C-17 Airbus 310 340 \\
    
        Amadori Jouannet and Krus \cite{AmadoriJouannetKrus2006} &	Catia & Matlab &	None & Unmanned Combat Aerial Vehicle \\
    
        Doyle et al \cite{Doyle} & Not mentioned & GMAP	& Plot3D
	   & High Speed Civil Transport and HWB N2A \\
    
        Cavallaro \cite{CavallaroDemasi2016}	& Catia &	Grid Generator	& GUI	& Aero-propulsion \\
    \bottomrule
    
    \end{tabular}
    }
    \label{tab:my_label}
\end{table}

\subsection{Novel contributions and outline of this paper}

In this paper, we present an integrated and automated environment to be used together with PanAir, a panel method based numerical modelling software developed by Boeing engineers in collaboration with NASA. PanAir, although being developed in early 70's is still in active use for the aircraft design. Since PanAir was developed as an in-house algorithm, the interaction with it can be tedious, especially in pre and post processing steps. In the pre-processing case, the user can take advantage of built in geometry discetization/panelling algorithms which comes together with PanAir; namely MakeWGS and Panin; however, these algorithms are not simply capable of accurately processing the aircraft geometry, and hence limiting the use of PanAir. Moreover, once the geometry is paneled, the user should manually check the orientation and connectivity with neighbouring panels; otherwise, PanAir can produce erroneous results. On top of that, the flow properties should be introduced in a specific format, otherwise, PanAir would not run analysis on the aircraft geometry of interest. The need for manual manipulation at several stages of preprocessing push researchers to develop either their own integrated environment or go through couple of more software to be able to create the required input geometry and define flow properties properly. 

The post-processing stage, similar to the pre-processing, requires manual manipulations of the output files of PanAir. The PanAir output format is not compatible to commonly used visualization tools, and hence, a translator algorithm is needed to pass the results to a visualization software. Moreover, if the results of the PanAir analysis are needed to be plotted, those plots should be created manually using a third party software or by mini algorithms written by the researcher, specific to his/her problem. 

In this study, we propose an advanced and comprehensive solution that addresses the limitations of existing approaches in creating a holistic, integrated, and automated environment for aerodynamic analysis. Unlike previous researchers such as Schmidt \cite{Schmidt}, Mehdi and Masud \cite{Mehdi}, Doyle et al. \cite{Doyle}, and Tarkian and Tessier \cite{TarkianTessier2008}, who utilized PanAir but faced difficulties in resolving abutment issues and establishing automatic software linkage, our approach overcomes these challenges. While Amadori, Jouannet, and Krus \cite{AmadoriJouannetKrus2006} took a step towards automation by completely automating the pre-processing  using Matlab, they did not address post-processing aspects. On the other hand, Cavallaro's \citep{CavallaroDemasi2016} work with a fully-automated PanAir environment showed promise, but it had limitations in importing Catia files, modeling the engine, and generating quadrilateral panels. Building upon these previous efforts, our research aims to create a fully-automated and integrated environment that streamlines the entire process. By taking the geometry coordinates as input, our system generates the output file within seconds, which is automatically exported to TecPlot for visualization. This integrated environment offers greater usability and efficiency compared to previous approaches, as it seamlessly handles both pre and post-processing tasks. With a focus on enhancing the automation and integration aspects, our proposed solution addresses the shortcomings of prior studies and provides a more robust framework for aerodynamic analysis. The use of appropriate vocabulary and sentence structure helps convey the significance and novelty of our research.

This paper is organized as follows. In Section \ref{sec:OldPanAir}, we have summarized the operational procedure of PanAir in the absence of an integrated environment, emphasizing the challenges involved. In the next Section \ref{sec:NewPanAir}, we have detailed the background processes running in proposed environment and emphasized the significantly reduced manual effort and substantial ease of use of PanAir when used together with the proposed environment here. Lastly in Section \ref{sec:numericalValidation}, we have tested the performance of the PanAir together with the integrated environment over a Cessna aircraft geometry. The flow around the aircraft is analyzed using PanAir within the integrated environment. The numerical results have been compared to wind tunnel test data, affirming not only their excellent agreement but also the significant reduction in analysis time from weeks of manual manipulations to mere minutes with the provided environment. The paper is closed with highlights, conclusions and future work presented in Section \ref{sec:conclusions}.

\section{Methodology and Workflow of PanAir} \label{sec:OldPanAir} 
\begin{figure}[ht]
  \begin{center}
    \includegraphics[width=0.90\textwidth]{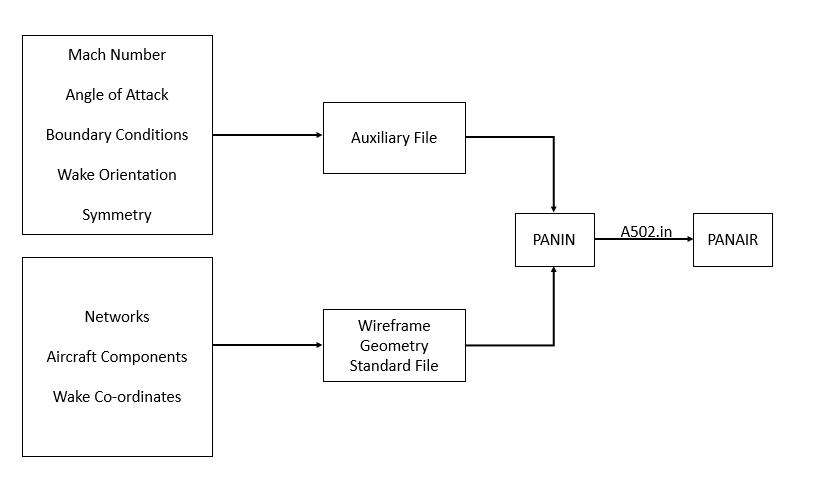}
\end{center}
\caption{A detailed flowchart describing the required files for PanAir and the information these files contain}
  \label{fig:PanAirWorkflow11}
\end{figure}

\begin{figure}[ht]
  \begin{center}
    \includegraphics[width=0.90\textwidth]{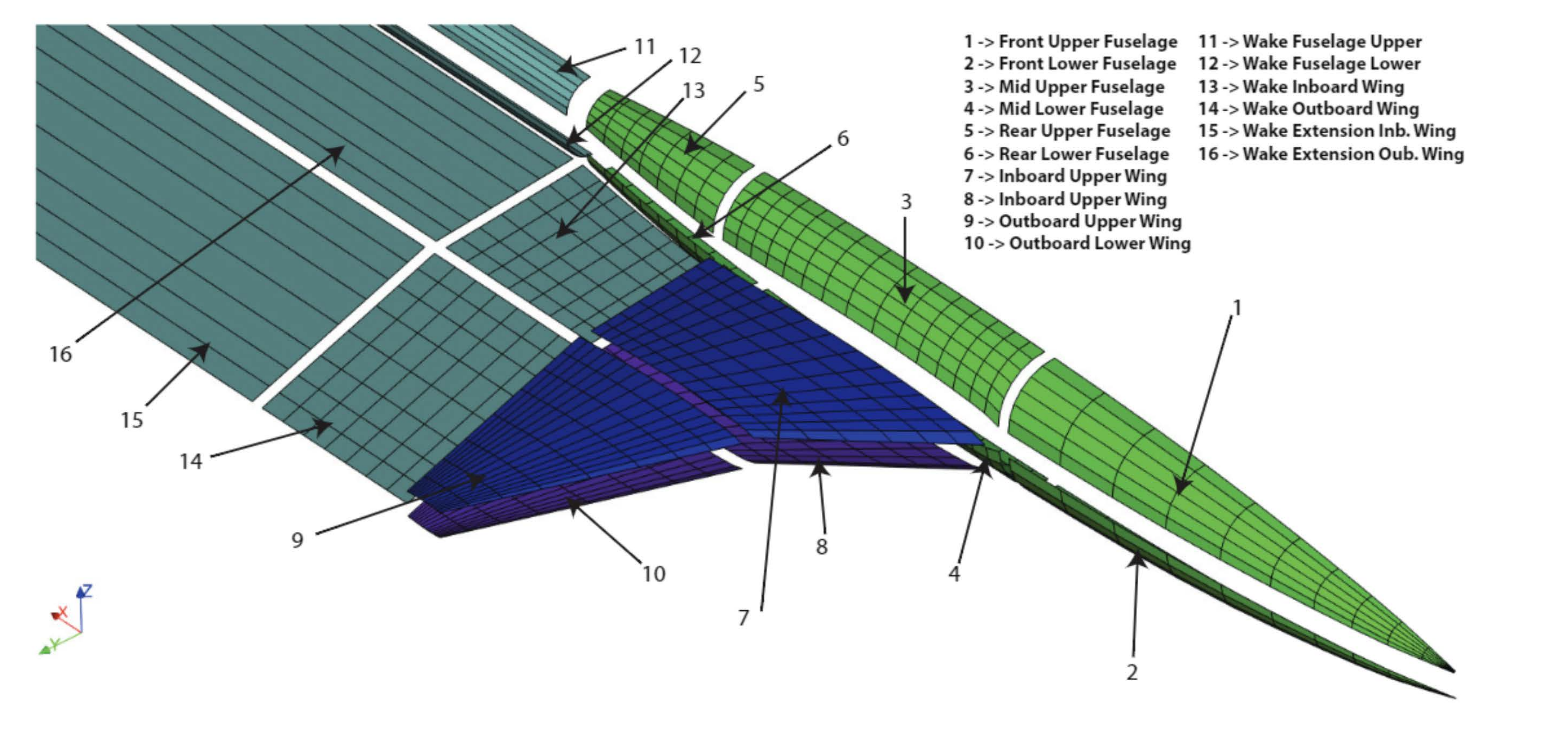}
\end{center}
\caption{Networks over a generic aircraft \cite{verstraetebsizing}}
  \label{fig:Network}
\end{figure}

In this section the standard PanAir workflow without any automated integrated environment used, is explained in detail, this workflow is schematically shown in Figure \ref{fig:PanAirWorkflow11}. To discretize the geometry in the standard PanAir implementation the built in pre-processors of PanAir, namely, MakeWGS and Panin can be used. 
The standard pre-processing has two steps, first, MakeWGS takes the geometry coordinates and discretizes the surface geometry into panels and networks. A network is a group of panels describing a certain aircraft component, like upper, lower, or central region, see Figure \ref{fig:Network}. The input data for MakeWGS is the leading-edge radius of the airfoil, the wing's span, and the fuselage's mean radius. The output format of MakeWGS is Langley Wireframe Geometry (LaWGS)which is the only compatible input format for the second step of preprocessing, that is executed using Panin.   
The LaWGS file obtained as an output from MakeWGS has the geometric data in quadrilateral panels written in a specific format. The other necessary input file for second preprocessing step is the auxiliary file; that is simply a text file with specific keywords (defined in PanAir Manual \cite{PanAir_Manual}) including all the flow data such as Mach number, angle of attack, and side-slip angle. Additionally, the auxiliary file contains boundary conditions specified for each network, wake coordinates at each lifting surface, their location and orientation, and other features for comprehensive analysis. Various boundary conditions can be imposed depending on the analysis. These boundary conditions include but are not limited to flow through impermeable surfaces, flow through inlets and fan faces or super-inclined surfaces. 


The LaWGS file and the auxiliary file are then combined by the second built-in pre-processor named Panin, and all the input data is structured into a particular format (6E10.0) with six numbers and ten spaces per line where data is divided into specific data blocks \cite{PanAir_Manual}. This structured input data created by Panin is crucial because PanAir is case-sensitive and keyword-specific software. The primary purpose of Panin is to only combine the data from the auxiliary file and LaWGS file and translate it into PanAir-compatible input format. It generates an input file which is a single input to PanAir and has all the necessary data to perform the analysis.
The PanAir output files consist of two key components: the ffmf file, which provides overall force coefficients, and the agps file, which contains the pressure coefficient for each node. However, the agps file, particularly with its nodal pressure coefficient variation, can be quite extensive and challenging to comprehend without an effective visualization tool. Without a dedicated post-processing tool, it becomes arduous to assess whether the right-hand rule is satisfied or if there are any issues with abutment or flow leakage. Hence, the availability of a suitable post-processing tool becomes crucial for accurately interpreting and analyzing the results, ensuring proper visualization, and facilitating the assessment of key factors like the adherence to the right-hand rule and the absence of any undesirable flow patterns.


\subsection{Shortcomings of the standard method}

As detailed in the previous section, use of PanAir includes many intermediary steps which makes the user interaction with the software difficult. Starting from the pre-processing, in this subsection, we have underlined the shortcomings of the standard PanAir use. As reader may remember, one of the two built in pre-precossor of PanAir is MakeWGS; that is used to discretize the geometry of interest. The  MakeWGS, however, can not handle complicated geometries, so alternative CAD software is often needed to discretize the geometry. There are other alternative third party, ready-to-use executable files available to model simple geometries into LaWGS format like Wingwgs but it can only model a generic rectangular wing. Similarly, other simple executables such as wd2wgs, wb2wgs, a5022wgs are all available which can translate a conventional wing-body geometry into LaWGS format required as PanAir input, but all comes as an extra step to overcome to get to the PanAir analysis. Most of the times, available software is not capable of discretizing the geometry of interest, and hence leading to erroneous lift and induced drag approximations \cite{Schmidt}. In this case the reseracher is left with no choice other than writing a mini translator algorithm specific to his/her problem. 

\begin{figure}[h!]
  \begin{center}
    \includegraphics[width=0.75\textwidth]{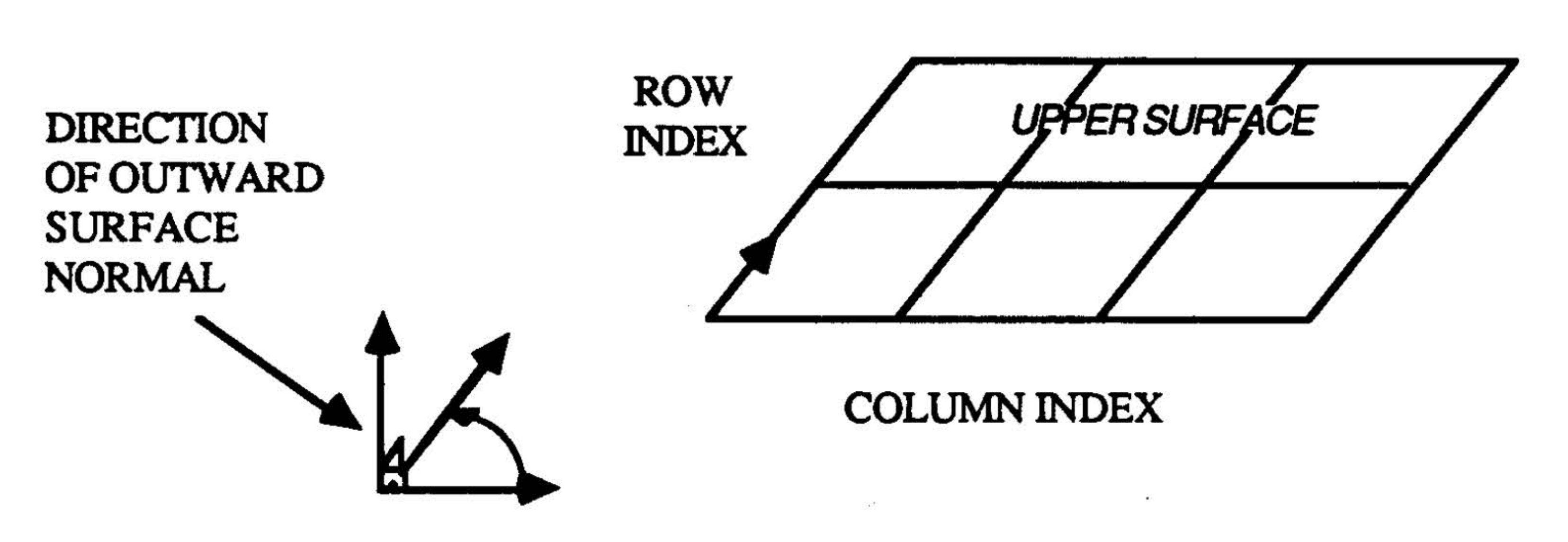}
\end{center}
\caption{The diagram representing right hand rule and how it is applied over the network to ensure flow tangency \cite{PanAir_Manual}}
  \label{fig:RHS}
\end{figure}

Thus, parametric modeling tools such as Catia and Gambit are readily being utilized for geometry discretization as evident from the literature review. The geometry file obtained from these tools, though, does not conform to the LaWGS format and therefore requires manual manipulations. Significant man-hours is required to ensure flow tangency throughout the discretized geometry obtained using Catia and Gambit. Moreover, the coordinates of each network panel should be manually arranged so that the surface normal unit vector point in the direction of the flow stream, see Figure \ref{fig:RHS}.

Manipulations are also required to ensure correct abutments; that is, each network is connected with the neighbouring one so that the flow does not leak into the body. It might look easy to conform, but by far this requirement is the most time-consuming. 

\begin{figure}[h]
\centering
\subfigure[]{
\includegraphics[width=0.75\textwidth]{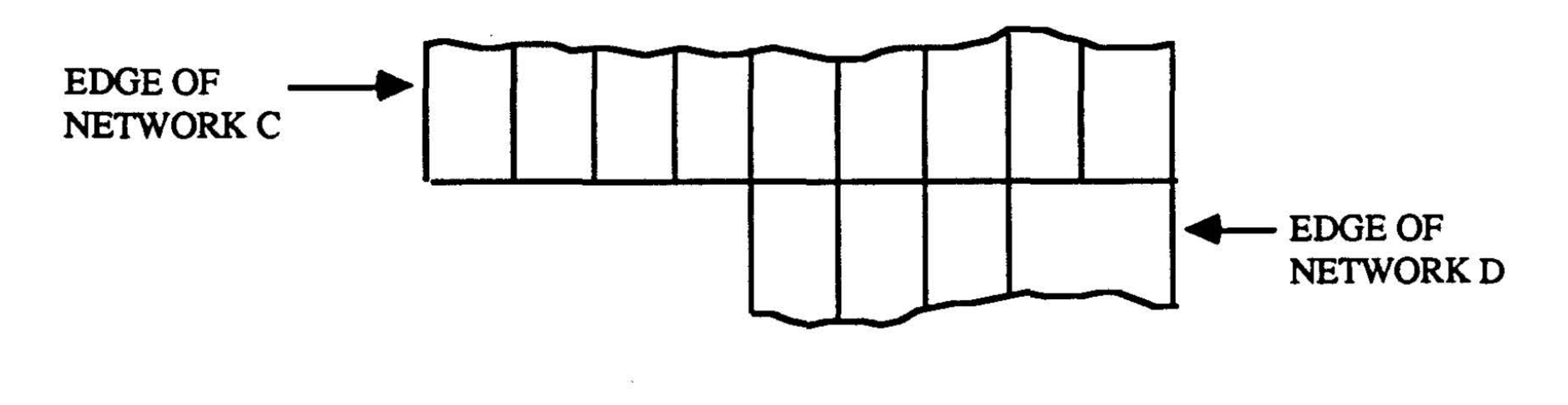}
}
\subfigure[]{
\includegraphics[width=0.75\textwidth]{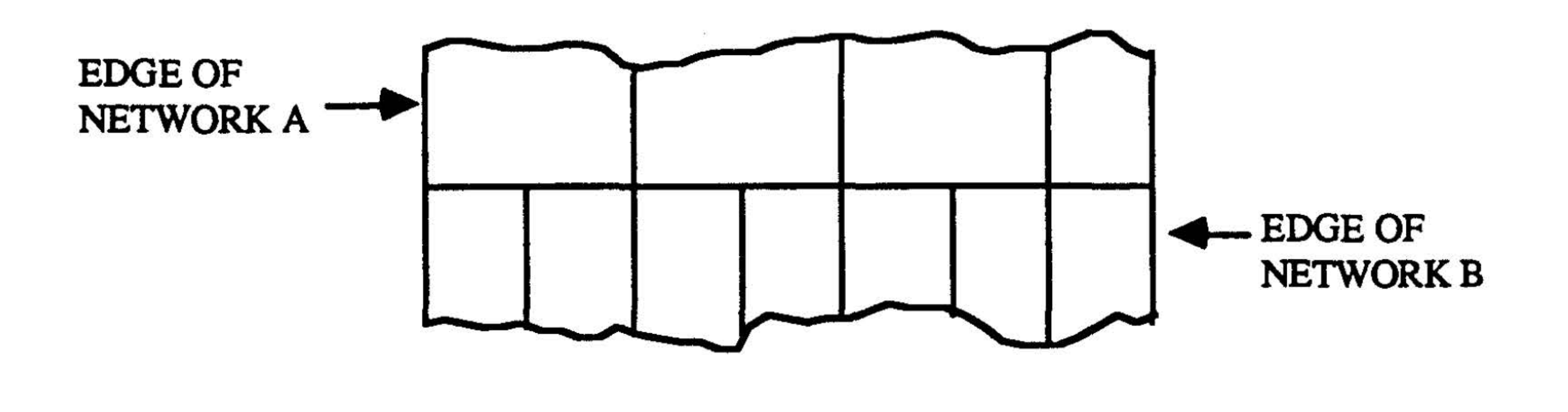}
}
\caption{Abutted and un-abutted edges between two networks \cite{PanAir_Manual}}
\label{fig:Abutment}
\end{figure}

As it is evident from the Figure \ref{fig:Abutment} the upper and lower network of the fuselage and wing must be connected to a common coordinate point. This requirement is mandatory for accurate flow simulations with PanAir as these un-abutted edges will distort the flow field and the flow will slip inside the geometry leading to erroneous results. This is done manually by replacing the edge point of the corner panel with the other network's edge point, which sometimes distorts the panel into curved shapes, that alters the geometry. 

After completing the preprocessing stage and running the analysis with PanAir, the subsequent challenge lies in effectively visualizing the obtained results. This task can be quite cumbersome, particularly due to the incompatibility of PanAir output files with commonly used visualization software. Consequently, manual manipulation of the output data, similar to the pre-processing stage, becomes necessary. The PanAir output files are simple text files, and in order to facilitate visualization, the agps file needs to be converted into the tecplot specified dat format. This conversion involves arranging the data according to the total number of nodes in each panel and the total number of panels in the network. By following this process, the output data can be suitably prepared for visualization, allowing for a more comprehensive and intuitive understanding of the results obtained from the PanAir analysis.
These laborious process necessitates a more efficient and streamlined approach to effectively interpret and extract meaningful insights from the data.


With the environment proposed in this work, we account for all these shortcomings and introduce a time-efficient and automatic pre-processing procedure. 

\section{Methodology and Work flow of PanAir with Proposed Environment} \label{sec:NewPanAir}

\begin{figure}
  \begin{center}
    \includegraphics[width=1.0\textwidth]{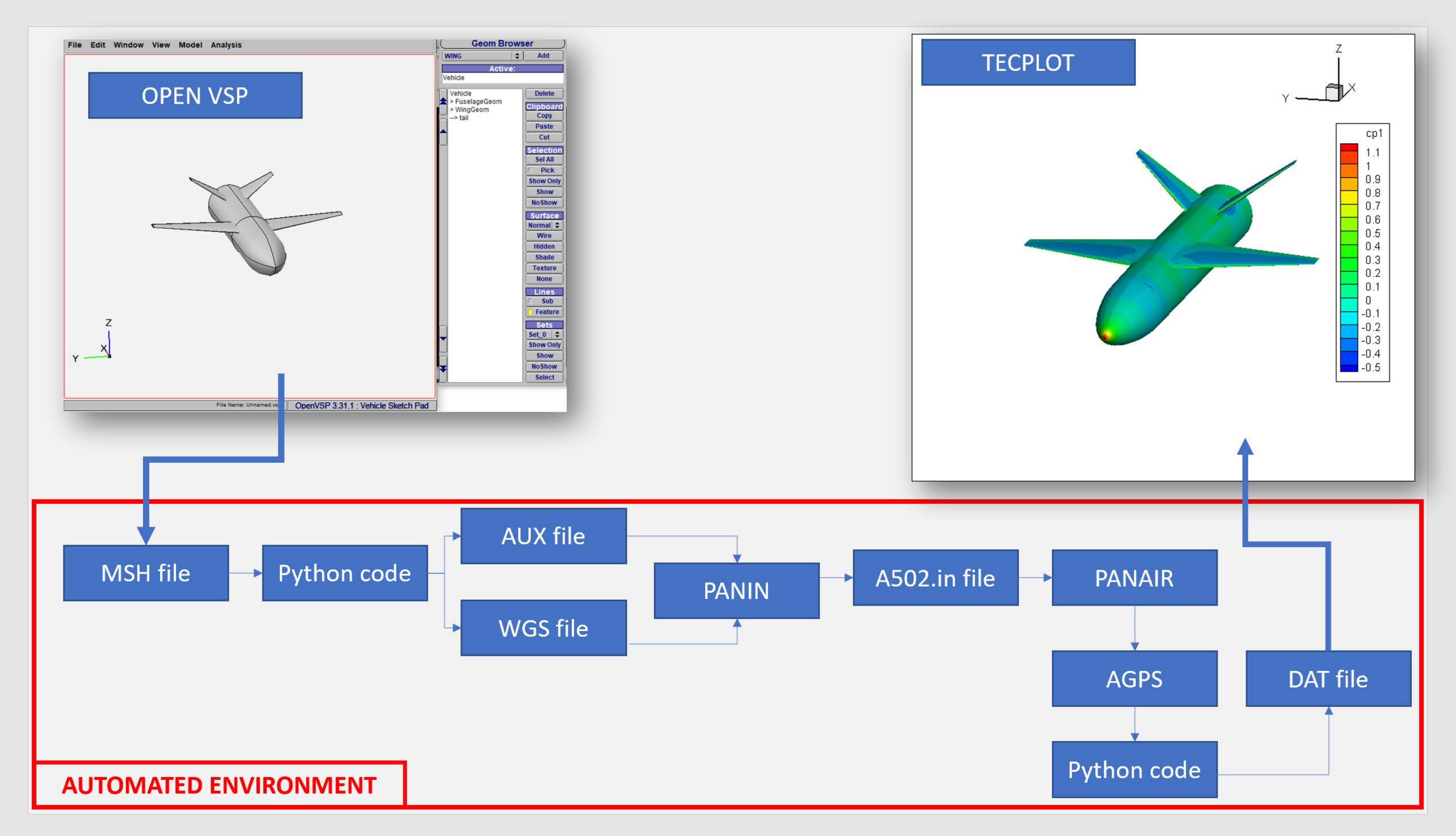}
\end{center}
\caption{A modified methodology for a PanAir integrated automated design environment}
  \label{fig:modflow}
   \end{figure}

Up until now, we have detailed how to use PanAir without any automated integrated environment and underlined the challenges when doing so. In this section, the details of the proposed automated integrated environment to PanAir is presented. The boundaries of integrated and automated environment is depicted schematically in Figure \ref{fig:modflow}. The integration process is specifically designed to seamlessly incorporate a .msh file, which is automatically obtained from OpenVSP. The reason for choosing OpenVSP is that the generated .msh file from this software aligns the geometric coordinates in the required format for the integrated environment to operate smoothly. By automating this step, the environment eliminates the need for manual handling of the .msh file. Other software such as Gambit and Catia also produces a .msh file which is in a different format so the same algorithms are not compatible with these files. Once the .msh file is imported into the environment, the preprocessing, analysis and postprocessing are efficiently carried out, culminating in the quick and convenient display of results within the Tecplot environment. This streamlined workflow significantly reduces manual intervention and accelerates the entire analysis, contributing to improved efficiency and ease of use. In this subsection, the background processes running in the integrated environment proposed in this work is explained in detail. From user's point of view, working with PanAir together with the integrated environment is as simply as inputing the .msh file and obtaining the visuals in the pop-up Tecplot window screen, leading to a user friendly smooth interface.  

To determine the aerodynamic characteristics of the aircraft, the first step is to model the geometry in any CAD software. OpenVSP geometry modeling module is widely used in both industry and academia by aerospace engineers \cite{kim2022openvsp}. It is relatively easier to use and, vital for present work, provides the advantage of having an integrated meshing tool that automatically defines each node of an aircraft in a global coordinate system. The geometry can also be imported from the online VSP Hangar where commonly used aircraft geometries are pre-defined. 

At first, the proposed environment takes as an input, the mesh file (.msh) for each aircraft component created by OpenVSP. User can set higher number of nodes at the leading edge of the wing as more accuracy is required to capture the significant pressure changes along the leading-edge radius. The concentration of nodes can be very low in less-featured regions. Each component file is then automatically sorted in accordance with the right-hand rule that guarantees the flow to be solved on the upper surface. The coordinates of each component are arranged in accordance with their cross-sections in the OpenVSP mesh file and also to conform to the right-hand rule. In order to ensure the surface normal vector points outward, the wing and horizontal tail points is sorted such that their upper surface points are organized sequentially from trailing edge to the leading edge and vice versa for their lower surfaces. Similarly, the fuselage points are automatically sorted from nose to tail in a clockwise manner. 

In order to simplify the sorting procedure, the aircraft components are divided into various networks. A network comprises of an ordered array of panels where the unit vector of each panel points in the same direction \cite{ShahidSajjad2020}. To make the computation easy and time-efficient, half configuration is modeled and the symmetry option is utilized. The sorted coordinates are then placed automatically in the LaWGS file as per the required format. Moreover, the correct abutment of networks are ensured within the integrated environment by letting the number of fuselage cross-sections in the vicinity of the wing to be equal to the number of camber points of the wing using spline approximation. A sixth-degree polynomial is modeled to represent the surface and to obtain other required nodes. All the sorting, abutting and file writing is done using a generic Python algorithm which can handle all the steps automatically. 

The user is then requested to input the flow conditions such as Mach number and angles of attack along with a boundary condition over each network together with the orientation of the wake. The Mach number must not be in transonic regime from 0.8 to 1.2 as the PanAir program is not able to evaluate the non-linear transonic behavior. Moreover, since the stall effects are not catered; the angle of attack is also limited to a maximum of 20 degrees. The integrated algorithm then invokes the PanAir built in preprocessor, Panin. Panin takes the auto-generated auxiliary and LaWGS files and readily generates the a502.in file which conforms to the PanAir input format.

The PanAir executable file is automatically invoked once the a502.in input file is generated. PanAir processes the file accordingly and generates multiple output files. Each output file contains different analysis-related data. The agps output file outlines the surface pressure across each node point for each angle of attack. Other files; namely ffm and ffmf; contain a summary of the resultant forces and moments on half geometry and over full geometry respectively. The lift and induced drag coefficients at respective angles of attack are listed in these files as well which can be sufficient to comprehend the results.

The results contained in agps output file can be visualized through a post-processing software such as Tecplot. Tecplot requires the data in the form of a specially-formatted text file called data (.dat) file. The developed environment converts the agps file into .dat file by performing successive translation operations including formatting specified title cards called headers for each network. A header contains name of the network and number of rows and columns with-in, all of which is written by the program developed using Python.
To obtain a Tecplot plot; certain commands such as geometry limit setting in the frame or obtaining pressure contours at each angle of attack are automated within the integrated environment. Tecplot provides an option to save the interactive commands in a macro (.mcr) file. This file is then included in the environment and after the Tecplot pops up successfully, the commands for visualizing results at each angle of attack, isometric and planform view of the aircraft and wake visualization are executed automatically.

\section{Numerical Validation} \label{sec:numericalValidation}

\begin{figure}[H]
  \begin{center}
    \includegraphics[width=0.75\textwidth]{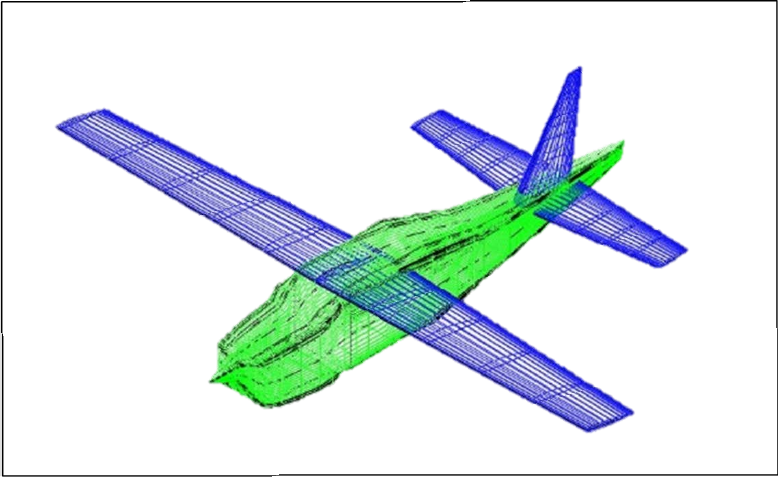}
\end{center}
\caption{Open-VSP model of Cessna 210 aircraft}
  \label{fig:cessna}
\end{figure}

To test and observe the performance of integrated environment for PanAir proposed in this work; Cessna Model 210 with modified airfoil of NLF(1)-0414F is analyzed using PanAir and the results are compared with wind tunnel experiments. The Cessna airfoil is chosen for this analysis since this airfoil belongs to a class of advanced airfoils specifically designed to maintain \textit{natural laminar flow} (NLF) and is best suited for subsonic flight regime. NLF airfoils theoretically proved to give maximum cruise performance, therefore, they were heavily tested and optimized at NASA Langley Research Center. A full-scale Cessna 210 model with slightly modified  NLF(1)-0414F wing is tested in the Langley 30 by 60 foot wind tunnel and the results made available in \cite{Windtunnel}. 
To get the numerical results using PanAir together with the proposed environment; the Cessna aircraft geometry, see Figure \ref{fig:cessna} is first created using Open-VSP using the geometry specifications presented in Table \ref{tab:spec}.
%
\begin{table}[ht]
    \centering
    \caption{Geometric data of modified Cessna NLF 210 utilized for analysis}
    \begin{tabular}{|c|c|}
\hline
Wing Aspect Ratio &	8.64 \\
\hline
Wing Taper Ratio &	0.65 \\
 \hline
Fineness ratio	& 6 \\
 \hline
Vertical tail area	& 13.53 ft2 \\
 \hline
Horizontal tail area &	47.59 ft2 \\
\hline
\end{tabular}
\label{tab:spec}
\end{table}
%
While paneling, at the wing and at the horizontal tail, finer panels -panels with smaller size- are defined since those parts are the major contributors to lift and induced drag. Fuselage is paneled homogeneously -no significant change in panel sizes- except where the wing and horizontal tail are attached. At those connection points finer panels are used to ensure correct abutments. 

\begin{figure}
  \begin{center}
    \includegraphics[width=0.75\textwidth]{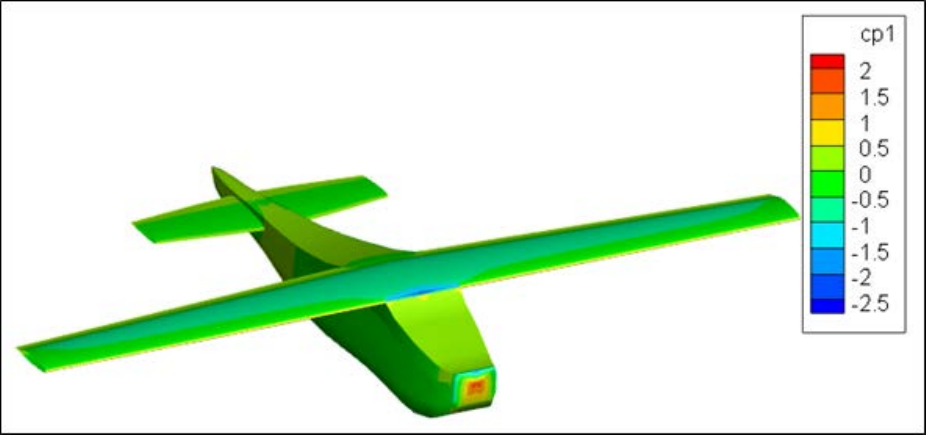}
\end{center}
\caption{Surface pressure contours at zero-degree angle of attack on Cessna 210 obtained in Tecplot}
  \label{fig:cont}
\end{figure}

After generating the geometry, the .msh file is simply fed into the integrated environment. Within couple of minutes (depending on the number of panels and complexity of the geometry) the PanAir analysis results pop-up in the screen. Engineers often interested in pressure distribution over the aircraft under certain flow conditions; as it aids in computing moments and forces over the aircraft. In Figure \ref{fig:cont} the pressure distribution of Cessna 210 at zero-degree angle of attack is presented. The flow is halted at the nose and the flow accelerates over the top surface causing lower pressure regions. On the lower surface, however, the flow decelerates and the pressure is high. Similar pressure gradients can be observed over the wing surfaces. Due to these gradients, lift is generated. The Tecplot screen in Figure \ref{fig:cont} showing the pressure distribution over the aircraft (or other outputs specified/interested) pops up automatically thanks to the integrated environment proposed here. For this analysis, the flow conditions used are stated in Table \ref{tab:sim}. 

Once the analysis is done, careful consideration is needed before comparing the PanAir results with the wind tunnel test data. PanAir only provides analysis for inviscid flows and it only provides the induced drag once the analysis ends. However, there is as well a so called parasite drag, that is significant in low velocity, high Reynolds number flows over all aircraft geometry. For a fair comparison with the wind tunnel data the parasite drag needs to be calculated and added into PanAir analysis. We have calculated the parasite drag using a mini Matlab algorithm, based upon Sommers and Shorts experimentally-determined equation \cite{Sommershort}. This equation is a modified version of \textit{reference temperature method} which determines skin friction drag for both laminar and turbulent flows. The skin friction drag coefficient is computed to be 0.035 for a velocity of 53.3 mph and is added to the induced drag coefficient for each angle of attack, for the comparison. Vertical tail of the aircraft has no contribution in lift and induced drag of the aircraft and hence, to simplify the analysis, vertical tail component is not taken into account. 

\begin{figure}[H]
  \begin{center}
    \includegraphics[width=0.90\textwidth]{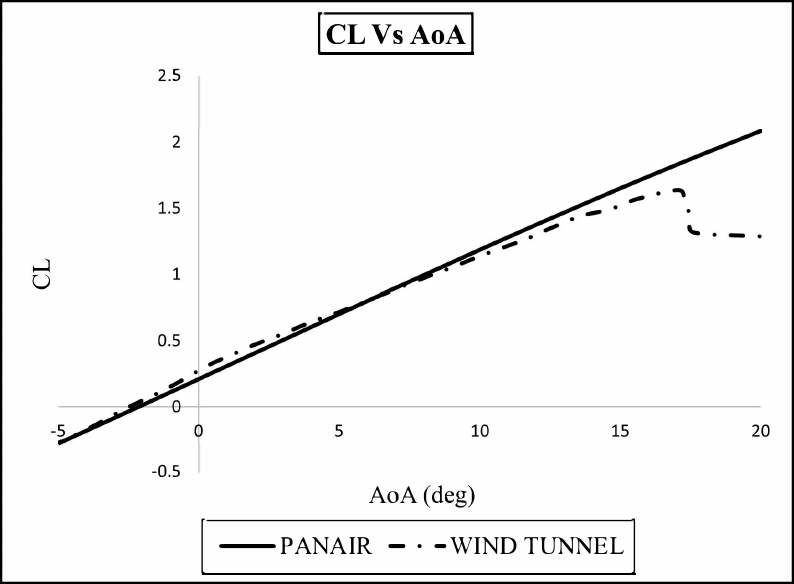}
\end{center}
\caption{Lift coefficient ($C_{L}$) Vs angle of attack (AoA) curves at M=0.07}
  \label{fig:cl}
\end{figure}

\begin{figure}
  \begin{center}
    \includegraphics[width=0.90\textwidth]{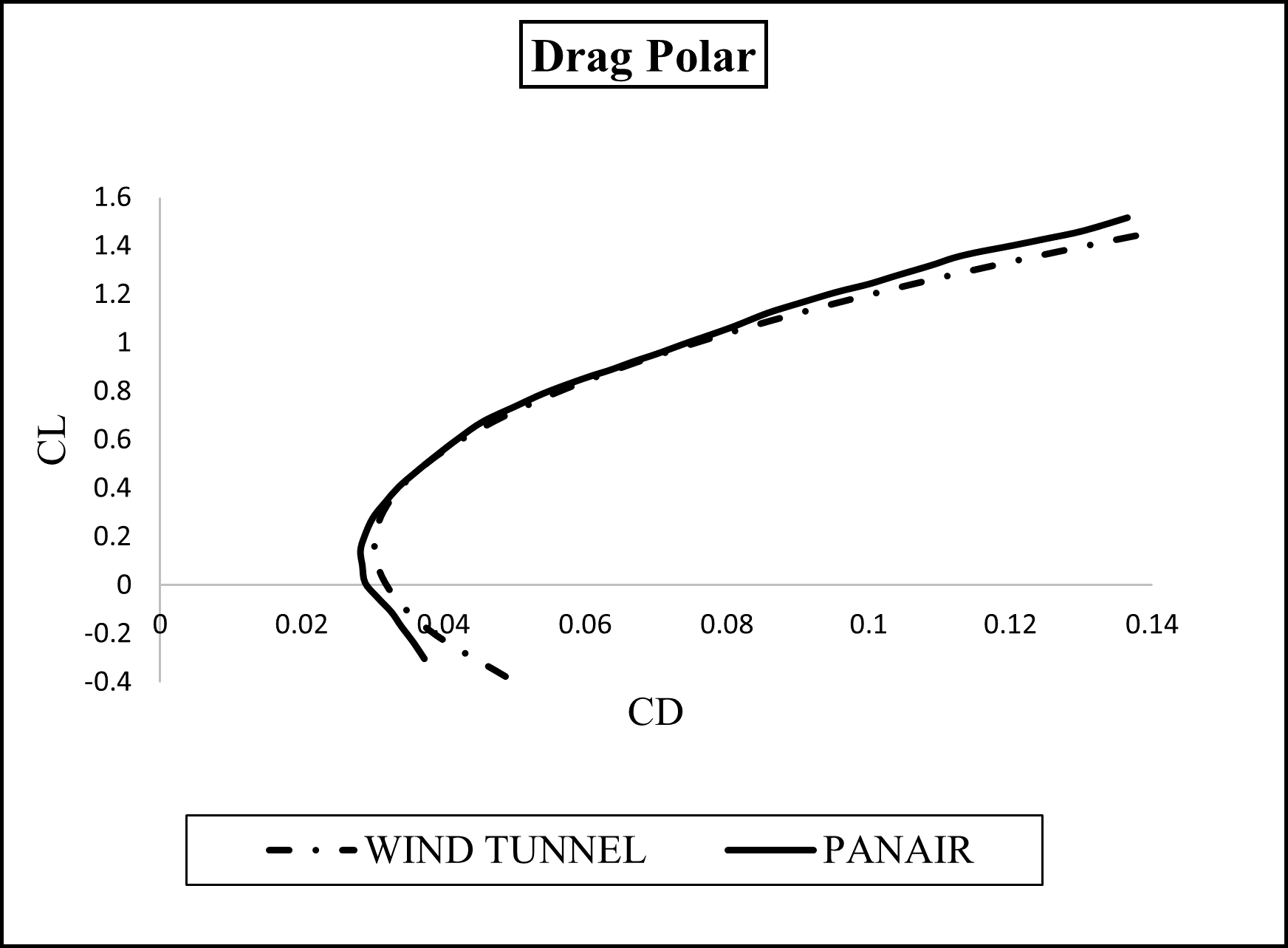}
\end{center}
\caption{Drag Polar at M=0.07 obtained from PanAir compared with wind tunnel results}
  \label{fig:cd}
\end{figure}

The results are presented in Figures \ref{fig:cl} and \ref{fig:cd}. In Figure \ref{fig:cl}; the lift coefficient ($C_{L}$) is calculated for different attack angles (AoA) and the results show excellent correspondence when compared with wind tunnel data especially when AoA is less than 15 degrees. The lift coefficient of an aircraft is dependent upon both the pressure and shear stress distribution. At this velocity -53.3 mph-, the shear stress contribution is insignificant as compared to the pressure contribution, hence, the viscous effects in lift can be ignored. At higher angles of attack, greater than 15 degrees, the results deviate from the experimental data. This is due to the fact that PanAir is based upon linearized potential flow theory i.e. any non-linearity in the flow is not catered for. Here, when AoA is greater than 15 degrees, we observe the non-linearity due to the stall effects that imply that the flow over the aircraft is no longer attached to the surface. This detachment creates adverse pressure gradients, causing the lift to decline ominously as shown by the wing tunnel data. These results are quite valuable in the sense that they are obtained in couple of minutes but still shows excellent agreement with very expensive wind tunnel tests.   
\begin{table}[ht]
    \centering
    \caption{Input Conditions for running PanAir}
    \begin{tabular}{|c|c|}
    \hline
         Velocity &	53.3 mph \\
         \hline
Reynolds Number	& $2x10^6$ \\
\hline
Number of panels	& 1613 \\
\hline
Total number of networks	& 9 \\
\hline
Total nodes	& 1938 \\
\hline
Altitude	& Sea-level \\
\hline
Angle of Attack &	0 to 20 degrees \\
\hline

    \end{tabular}
    
    \label{tab:sim}
\end{table}
\begin{figure}[H]
  \begin{center}
    \includegraphics[width=0.75\textwidth]{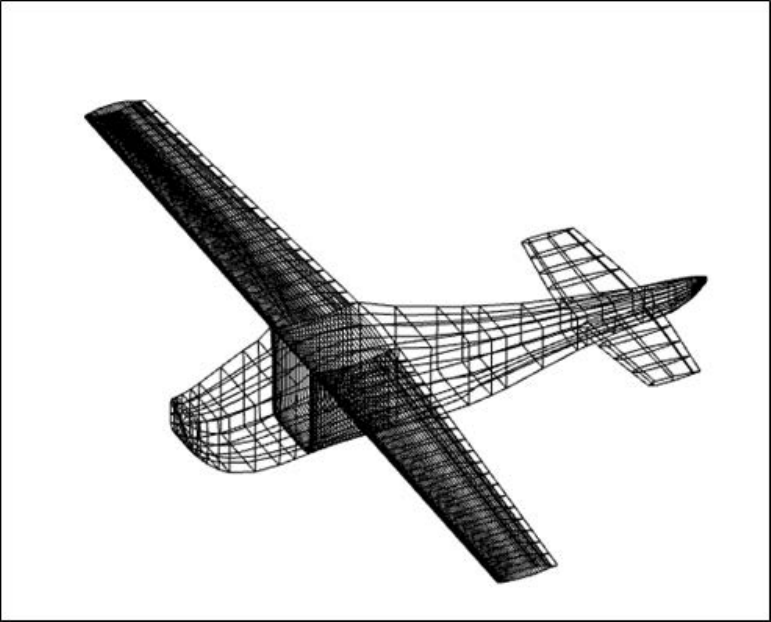}
\end{center}
\caption{Mesh generated with-in the integrated environment}
  \label{fig:PanAirWorkflow}
\end{figure}

Drag polar shown in Figure \ref{fig:cd} shows how drag coefficient changes with respect to the lift coefficient. The drag polar is one of the major contributors in determining the performance of any aircraft. It is often required to minimize the drag force and maximize the lift force throughout the flight regime but due to various factors, this condition cannot be maintained practically. Subsonic aircraft performance is, therefore, only optimized for the cruise condition. As it is presented in Figure \ref{fig:cd} the drag polar obtained after PanAir analysis is in great agreement with wind tunnel test data, proving the performance of PanAir in aircraft analysis.

\section{Conclusions}\label{sec:conclusions}
We have presented an integrated environment to be used together with PanAir, a high-order aerodynamic panel method-based software developed as a part of the Public Domain Aeronautical Software program with
NASA sponsorship. PanAir is extensively used to analyse the flow over aircraft to mainly calculate the forces acting over the wing and the body. Being an initially-in-house developed algorithm, PanAir does not offer a user friendly interface; instead, the user needs to manually manipulate input file or write enhancing algorithms specific to their problem, to be able to run PanAir analysis on their aircraft of interest. This requires significant man-hour, extending the time to run the analysis to days or even to weeks. 

As a possible remedy to this problem, here we have presented an integrated environment to be used together with PanAir. The user simply needs to feed the integrated environment with the geometry/mesh data and the PanAir analysis results pops up in the screen; overstepping several intermediary steps necessary otherwise. With the integrated environment, any aircraft geometry can be analysed using PanAir (otherwise not possible); removing the need for developing problem-specific enhancing algorithms. 

The current difficulties of using PanAir is underlined; after that, the underlying algorithms behind the proposed integrated and automated environment are presented. Numerical results prove that the integrated environment significantly improve the user interaction with PanAir. Moreover, results obtained using PanAir together with the integrated environment shows perfect agreement with wind tunnel experiment data.         
\section{Acknowledgements}
This work is supported by The Scientific and Technological Research Council of T\"urkiye (T\"UBITAK), Career Development Program (CAREER), project no:121M947.  

\newpage
\bibliographystyle {siam}
\bibliography{biblio}

\end{document}